\documentclass[sigconf,natbib=true,screen=true]{acmart}

\usepackage{acronym}
\usepackage[skip=0pt]{caption}
\usepackage[inline]{enumitem}
\usepackage{enumitem}
\usepackage{booktabs} 
\usepackage{multirow}
\usepackage{array}
\usepackage{pifont}
\usepackage{xspace}
\usepackage{subcaption}
\usepackage{makecell}
\usepackage{CJKutf8}
\usepackage{tcolorbox}
\usepackage[para]{footmisc}
\usepackage{amsfonts}
\usepackage{rotating}
\AtBeginDocument{%
  \providecommand\BibTeX{{%
    \normalfont B\kern-0.5em{\scshape i\kern-0.25em b}\kern-0.8em\TeX}}}

\copyrightyear{2025} 
\acmYear{2025} 
\setcopyright{acmlicensed}\acmConference[SIGIR '25]{Proceedings of the 48th International ACM SIGIR Conference on Research and Development in Information Retrieval}{July 13--18, 2025}{Padua, Italy}
\acmBooktitle{Proceedings of the 48th International ACM SIGIR Conference on Research and Development in Information Retrieval (SIGIR '25), July 13--18, 2025, Padua, Italy}
\acmDOI{10.1145/3726302.3731686}
\acmISBN{979-8-4007-1592-1/2025/07}

\acrodef{AI}{artificial intelligence}
\acrodef{IR}{information retrieval}
\acrodef{LLM}{large language model}
\acrodef{CIS}{conversational information seeking}
\acrodef{CNP}{clarification need prediction}
\acrodef{CRF}{conditional random field}
\acrodef{HRL}{hierarchical reinforcement learning}
\acrodef{MLE}{maximum likelihood estimation}
\acrodef{MLP}{multilayer perceptron}
\acrodef{QPP}{query performance prediction}
\acrodef{RL}{reinforcement learning}
\acrodef{SAP}{system action prediction}
\acrodef{SIP}{system initiative prediction}
\acrodef{WISE}{wizard of search engine}
\acrodef{EEwRJ}{effectiveness evaluation without relevance judgments}
\acrodef{RAG}{retrieval-augmented generation}
\acrodef{NLP}{natural language processing}
\acrodef{PEFT}{parameter-efficient fine-tuning}
\acrodef{ICL}{in-context learning}
\acrodef{LoRA}{low-rank adaptation}
\acrodef{RR}{reciprocal rank}
\acrodef{AP}{Average Precision}
\acrodef{nDCG}{normalized discounted cumulative gain}
\acrodef{HSD}{Tukey's honestly significant difference}
\acrodef{ANOVA}{analysis of variance}
\acrodef{UEF}{utility estimation framework}
\acrodef{QPP-PRP}{pairwise rank preference-based QPP}
\acrodef{M-QPPF}{multi-task query performance prediction framework}
\acrodef{WRIG}{weighted relative information gain-based model}
\acrodef{ANCE}{Approximate nearest neighbor Negative Contrastive Estimation}
\acrodef{QPP4CS}{query performance prediction for conversational search}
\acrodef{CS}{conversational search}
\acrodef{CAsT}{conversational assistance track}
\acrodef{DCG}{discounted cumulative gain}
\acrodef{EET}{efficiency-effectiveness trade-off}
\acrodef{RDE}{re-ranking depth estimation}
\acrodef{CEC}{certified error control}
\acrodef{VDP}{variable-depth pooling}
\acrodef{RLT}{ranked list truncation}
\acrodef{LtR}{learning-to-rank}
\acrodef{RAML}{reward augmented maximum likelihood}
\acrodef{LSTM}{long short-term memory}

\newcommand{\todo}[1]{\textcolor{blue}{#1}}

\begin{document}

\title[Conversational Search: From Fundamentals to Frontiers in the LLM Era]{Conversational Search: \\ From Fundamentals to Frontiers in the LLM Era}

\author{Fengran Mo}
\orcid{0000-0002-0838-6994}
\affiliation{%
  \institution{Université de Montréal}
  \city{Montréal}
  \country{Canada}
}
\email{fengran.mo@umontreal.ca}

\author{Chuan Meng}
\orcid{0000-0002-1434-7596}
\affiliation{%
  \institution{University of Amsterdam}
  \city{Amsterdam}
  \country{The Netherlands}
}
\email{c.meng@uva.nl}

\author{Mohammad Aliannejadi}
\orcid{0000-0002-9447-4172}
\affiliation{%
  \institution{University of Amsterdam}
  \city{Amsterdam}
  \country{The Netherlands}
}
\email{m.aliannejadi@uva.nl}

\author{Jian-Yun Nie}
\orcid{0000-0003-1556-3335}
\affiliation{%
  \institution{Université de Montréal}
  \city{Montréal}
  \country{Canada}
}
\email{nie@iro.umontreal.ca}

\renewcommand{\shortauthors}{Fengran Mo, Chuan Meng, Mohammad Aliannejadi, and Jian-Yun Nie}

\begin{abstract}
Conversational search enables multi-turn interactions between users and systems to fulfill users' complex information needs. 
During this interaction, the system should understand the users' search intent within the conversational context and then return the relevant information through a flexible, dialogue-based interface.
The recent powerful large language models (LLMs) with capacities of instruction following, content generation, and reasoning, attract significant attention and advancements, providing new opportunities and challenges for building up intelligent conversational search systems.
This tutorial aims to introduce the connection between fundamentals and the emerging topics revolutionized by LLMs in the context of conversational search.
It is designed for students, researchers, and practitioners from both academia and industry.
Participants will gain a comprehensive understanding of both the core principles and cutting-edge developments driven by LLMs in conversational search, equipping them with the knowledge needed to contribute to the development of next-generation conversational search systems.
\end{abstract}

\ccsdesc[500]{Information systems~Users and interactive retrieval}
\ccsdesc[500]{Computing methodologies~Discourse, dialogue and pragmatics}

\keywords{Conversational information retrieval, Conversational search, Retrieval-augmented generation, Large language models}

\begin{abstract}
Conversational search enables multi-turn interactions between users and systems to fulfill users' complex information needs. 
During this interaction, the system should understand the users' search intent within the conversational context and then return the relevant information through a flexible, dialogue-based interface.
The recent powerful large language models (LLMs) with capacities of instruction following, content generation, and reasoning, attract significant attention and advancements, providing new opportunities and challenges for building up intelligent conversational search systems.
This tutorial aims to introduce the connection between fundamentals and the emerging topics revolutionized by LLMs in the context of conversational search.
It is designed for students, researchers, and practitioners from both academia and industry.
Participants will gain a comprehensive understanding of both the core principles and cutting-edge developments driven by LLMs in conversational search, equipping them with the knowledge needed to contribute to the development of next-generation conversational search systems.
\end{abstract}

\maketitle

\section{Motivation}
Search engines are essential for modern society to fulfill users’ information needs~\cite{white2025information}.
The initial ad-hoc search systems rely on keywords or short phrases as input queries, which cannot capture user search intent accurately with limited interactions~\citep{gao2022neural,zamani2023conversational,mo2024survey}.
The recently developed conversational search~\citep{yu2021few,mo2023convgqr,mao2023large,mo2024chiq,mo2024aligning} enables multi-turn interactions between users and systems to meet users' complex information needs, supporting the flexibility of interactions.
To further enhance interaction quality, conversational search systems are increasingly adopting mixed-initiative strategies, where both the user and the system can take the initiative at different points in a conversation~\cite{radlinski2017theoretical}.
Mixed-initiative enables the system to take initiative when appropriate~\citep{meng2023system}, such as by asking clarifying questions in response to ambiguous queries~\citep{aliannejadi2021building}, or by proactively providing background information or suggestions~\citep{meng2025bridging}, ultimately leading to a more satisfying user experience.

In the current era of large language models (LLMs)~\cite{zhao2023survey,zhu2023large, askari2024generative},  powerful language models further improve the capacity of understanding conversational context and complex queries. 
Besides, users are more used to interacting with systems in multiple turns.
The flexibility of the conversation interface provides more opportunities for conversational interactions, especially based on the integration between the search engine and LLMs~\cite{su2024dragin,jin2025search,li2025search}. 
For example, the input could be the context-dependent query, task-oriented instruction, multi-modal information, etc. 
Conversely, the output of the system could be in different forms -- direct answer, search result page, summarization of the search results, clarification question, and so on.
The various interactive modes of conversational search systems result in a possible paradigm shift of user information-seeking behavior -- their preference from single-turn to multi-turn conversation.
This ongoing shift underlies the potential of conversational search as a next-generation information-seeking system with new, broader opportunities and challenges. 

The build-up of a conversational system involves the integration of various components and techniques corresponding to different goals.
For example, the conversation modeling, clarification requirement detection, and response generation should depend on a specific design with corresponding knowledge, whose individual performance would influence the whole workflow.
Besides, the development of LLMs provides more possible techniques and opportunities to improve the systems, e.g., self-reflect, accurate relevance identification via reasoning, iterative actions planning, and retrieval-augmented response generation.

To the best of our knowledge, the latest tutorial~\cite{dalton2022conversational} on conversational search was presented at SIGIR 2022, which was before the emergence of the LLMs. 
Despite the existence of books~\cite{gao2022neural,zamani2023conversational,white2025information}, surveys~\cite{keyvan2022approach,zhu2023large,deng2025proactive}, and a large body of research~\cite{mo2023learning,mao2023learning,zhang2024usimagent,deng2024towards} on conversational search, these materials do not systematically introduce the connection between fundamentals and the emerging topics revolutionized by LLMs in the context of conversational search.
Therefore, we believe that our tutorial will present novel points of view, useful to the researchers and practitioners in conversational search, providing a clear view of the topic's evolution with the rise of LLMs.

\section{Objectives}
Conversational search is a rapidly growing field, with new concepts and techniques being continuously developed. 
It is also a multi-discipline area, whose development requires contributions from multiple communities, e.g., information retrieval (IR), natural language processing (NLP), human-computer interaction (HCI), speech, and dialogue.
With the recent advancement of LLMs, more potential integration mechanisms and applications are offered but are still underexplored, calling for more research.

This tutorial is organized around two main components in conversational search: (i) a survey-style overview of fundamentals and (ii) a deep dive into emerging topics shaped by the era of LLMs.
In doing so, we aim to provide newcomers with a solid grounding in conversational search while offering fresh perspectives and insights to students, researchers, and practitioners who are already familiar with the field and seeking to explore its future directions.
Concretely, this tutorial targets three primary objectives: 
\begin{itemize}[leftmargin=*,nosep]
    \item Provide a comprehensive overview of foundational topics in conversational search based on prior research.
    \item Present an up-to-date review of available datasets, evaluation protocols, and real-world applications.
    \item Explore emerging topics, including opportunities and challenges in developing conversational search systems with LLMs. 
\end{itemize}

\section{Relevance}

Conversational search is an emerging topic in the core IR communities by enabling the search in an interactive manner. 
It is also attractive in both industry and academia, and it is emerging as the new search paradigm~\cite{gao2022neural}.
Besides, the TREC Conversational Assistance Track (CAsT)~\citep{dalton2019cast,dalton2020cast,dalton2021cast,owoicho2022trec} has been organized by TREC since 2019, which evolved to the TREC Interactive Knowledge Assistance Track (iKAT)~\citep{aliannejadi2024trec,abbasiantaeb2025conversational} from 2023 to adapt the development of AI techniques.
This tutorial aims to introduce the recent advances in conversational search with the development of LLMs for the increasing interest of SIGIR audiences.

Several tutorials on the related topic have been presented in IR communities in recent
years, including 
\begin{itemize}[leftmargin=*,nosep]
    \item Recent Advances in Conversational Information Retrieval at SIGIR 2020 by~\citet{gao2020recent}.
    \item Interactive Information Retrieval: Models, Algorithms, and Evaluation at SIGIR 2021 by~\citet{zhai2020interactive}.
    \item Conversational Information Seeking: Theory and Evaluation at CHIIR 2022 by~\citet{aliannejadi2022conversational}.
    \item Conversational Information Seeking: Theory and Application at SIGIR 2022 by~\citet{dalton2022conversational}.
\end{itemize}

Different from the previous tutorials before the development of LLMs, we will put more focus on introducing the connection between fundamentals and emerging topics in conversational search based on LLMs.
Although some previous tutorials~\citep{liao2023proactive2,deng2023rethinking} have discussed the use of conversational agents in the era of LLMs, they primarily focus on general proactive dialogue systems (e.g., for task completion) rather than on search-oriented systems.

\section{Format and Schedule}
This tutorial is planned as a half-day event (3 hours), consisting of two 90-minute sessions with a break in between. It will be held in person. The proposed content will cover the following topics:

\vspace{0.50\baselineskip}
\noindent \textbf{Part I: Fundamentals of Conversational Search (90 minutes)}
\vspace{0.50\baselineskip}

\noindent \textbf{Introduction to Conversational Search [10 min]}.
We will begin this session by defining conversational search.
In particular, we will highlight the key characteristics that distinguish conversational search from traditional ad-hoc search.
For example, user queries in conversational search are context-dependent user queries, i.e., they may contain omissions, co-references, or ambiguities~\citep{yu2021few}, making it challenging for ad-hoc search methods to capture the underlying information need~\citep{radlinski2017theoretical}.
Moreover, we will explore real-world applications that demonstrate the growing importance and impact of conversational search in practice.

\noindent \textbf{Datasets and Evaluation [10 min]}. 
We will provide a quick overview of widely used datasets in conversational search.
For instance, OR-QuAC~\citep{qu2020open}, QReCC~\citep{anantha2021open}, and TREC CAsT 2019–2022~\citep{dalton2019cast,dalton2020cast,dalton2021cast,owoicho2022trec} are characterized by providing self-contained human-rewritten queries at each turn.
We will also cover TopiOCQA~\citep{adlakha2022topiocqa}, which introduces topic shifting, as well as the recent TREC iKAT 2023–2024~\citep{aliannejadi2024trec,abbasiantaeb2025conversational}, which incorporates user personal information into conversational search.
Regarding evaluation, we will introduce the metrics widely used in conversational search, such as traditional nDCG@3~\citep{dalton2019cast,dalton2020cast,dalton2021cast,owoicho2022trec} and recent nugget-based evaluation~\citep{abbasiantaeb2025conversational}.

\noindent \textbf{Conversational Search Paradigms [35 min]}.
Recovering the underlying information need from conversational history is a key challenge in conversational search~\citep{mao2022curriculum}.
To address it, two main approaches of conversational search methods have been developed: query-rewriting-based retrieval and conversational dense retrieval.
\begin{itemize}[leftmargin=*,nosep]
\item  Query-rewriting-based retrieval methods first rewrite a query that is part of a conversation into a self-contained query, which is then reused by an ad-hoc retriever~\citep{voskarides2020query,yu2020few,vakulenko2021comparison,lin2021multi,wu2022conqrr}.
Query rewriting can be done by either query expansion or sequence generation.
The former adds terms from the conversational history to the current query, e.g., by training a binary term classifier~\citep{vakulenko2021comparison}, while the latter directly generates the reformulated queries using pre-trained language models~\citep{yu2020few,lin2021multi}.
\item Conversational dense retrieval methods train a query encoder to encode the current query within the conversational history into a contextualized query embedding; the contextualized query embedding is expected to better capture the information need of the current query in a latent  space~\citep{qu2020open,yu2021few,lin2021contextualized,mao2022curriculum,mao2022convtrans}.
\end{itemize}

\noindent \textbf{Mixed Initiatives [35 min]}.
We will cover two core aspects of research on mixed-initiative interactions: \textit{what type of initiative to take} and \textit{when to take it}.
\begin{itemize}[leftmargin=*,nosep]
\item What type of initiative to take: clarification~\citep{aliannejadi2019asking,aliannejadi2021building,Hamedwww2020}, elicit user preferences~\citep{radlinski2019coached}, ask for feedback~\citep{trippas2020towards}, proactively provide background information/suggestions~\citep{meng2025bridging}, and so on. 
\item When to take the initiative: clarification need prediction~\citep{lu2025zero}, system initiative prediction~\citep{meng2023system}, and so on.
\end{itemize}

\vspace{0.50\baselineskip}
\noindent \textbf{Part II: Emerging Topics in the LLM Era (90 minutes)}
\vspace{0.50\baselineskip}

\noindent \textbf{Automatic Evaluation for Conversational Search [20 min]}.
We will discuss how to automatically evaluate the ranking quality of conversational search systems.
Recent studies explore two main approaches: query performance prediction (QPP)~\citep{meng2025qpp} and LLMs to generate relevance judgments.
\begin{itemize}[leftmargin=*,nosep]
\item  QPP for conversational search~\citep{meng2023query,meng2024query,meng2023performance,meng2025query}.
Existing QPP studies have shown that traditional methods~\citep{meng2023query} and the new proposed approaches specifically designed for the conversational setting~\citep{faggioli2023geometric,meng2023performance} yield promising results in conversational search.
\item  Using LLMs to generate relevance judgments for conversational search~\citep{meng2025query,abbasiantaeb2024can,abbasiantaeb2025improving}.
Existing studies either prompt LLMs~\citep{abbasiantaeb2024can,abbasiantaeb2025improving} or fine-tune them~\citep{meng2025query} to automatically generate relevance judgments, achieving high agreement with human annotators.
\end{itemize}

\noindent \textbf{Conversational Search with LLM-based Generation [30 min]}.
Conversational search based on the generative LLMs aims to integrate conversational search and LLMs. 
With the development of LLMs, conversational interfaces have become common in the interactions between users and systems. This changes the way to access information, and it generates new information needs and information access tasks, which heavily rely on the integration of search models and generative LLMs in an interactive manner~\citep{gong2024cosearchagent,mo2024survey}.
Two potential directions are emerging: (i) generation-augmented retrieval (GAR) and (ii) retrieval-augmented generation (RAG)~\citep{zhu2023large}.
\begin{itemize}[leftmargin=*,nosep]
\item Conversational generation-augmented retrieval (GAR)~\citep{mo2024convsdg}:  GAR can enhance the search systems by generating pseudo data, producing fluent and consistent interaction, identifying the information value among the generation, and decomposing complex or instruction-based queries.
\item Conversational retrieval-augmented generation (RAG): Similarly, the conversational RAG~\citep {cheng2024coral,katsis2025mtrag}, which integrates conversational search results into the prompt that can contribute to improving truthfulness in the answer generation of LLMs by grounding answers in the retrieved documents, thus reducing hallucinations. It also helps increase information diversity.
\item Collaboration between retrieval and generation: The collaboration is generally expected, but the concrete solutions in conversational search are still under-explored~\citep{su2024dragin,wu2025retrieval}, e.g., to determine the right time and way to do it. We will introduce different potential and promising directions in this context.
\end{itemize}

\noindent \textbf{Personalized Conversational Search [15 min]}.
The development of personalized conversational search systems has been driven by advancements in LLMs, aiming to tailor the generated outputs according to their users' background and historical preferences.
The additional challenge lies in dynamically incorporating users' profiles, preferences, or history into each turn~\citep{aliannejadi2024trec,mo2024leverage,abbasiantaeb2025conversational}.
We will introduce the recent advances in personalized conversational search. 

\noindent \textbf{Agentic Conversational Search [10 min]}. Compared with traditional conversational search, the agentic conversational search systems have the following advantages: (i) the systems are expected to not only satisfy users' information needs via search, but also complete users' information tasks via actions and interactions of LLM-based agents; (ii) the systems can improve the mixed-initiative levels by automatically determining the suitable action for each query turn~\citep{meng2023system}, and dynamically perform search via intelligent planning until satisfying the users~\citep{meng2024ranked}. 
We will discuss the possible directions of agentic/future conversational search systems.

\noindent \textbf{Summary and Discussion with the Audience [15 min]}.
In the final part of the tutorial, we will summarize the key topics covered in this tutorial and outline promising future research directions for conversational search in the era of LLMs. 
We will then open the floor for an interactive discussion with the audience to exchange ideas, insights, and questions.

\section{Materials}
%
To enhance the learning experience, we provide a comprehensive set of supporting materials in preparation for our tutorial. These resources will be made publicly available through a dedicated GitHub repository and Website, which will be accessible several weeks prior to the conference. This early access will allow participants ample time to explore and familiarize themselves with the content. 
Besides, some content is partly supported by a recent survey written by some of the oragnizers~\cite{mo2024survey}.

\begin{acks}
    This research was partially funded by Ahold Delhaize, through AIRLab Amsterdam, and a discovery grant from the NSERC.
\end{acks}

\bibliographystyle{ACM-Reference-Format}
\balance
\bibliography{references}


\end{document}